\documentclass[sigconf]{acmart}
\usepackage[most]{tcolorbox}
\usepackage[table,xcdraw]{xcolor}
\usepackage{booktabs}
\usepackage{pifont}
\usepackage{array}
\usepackage{balance}
\usepackage{adjustbox}
\usepackage{siunitx}

\newcommand{\syful}[1]{\textcolor{black}{#1}}

\setcopyright{cc}
\copyrightyear{2026}
\acmYear{2026}
\acmConference[EASE 2026]{The 30th International Conference on Evaluation and Assessment in Software Engineering}{9–12 June, 2026}{Glasgow, Scotland, United Kingdom}

\newcommand{\RqOne}{\texorpdfstring{RQ$_1$}: To what extent can we replicate the results of \citet{reis2023security} about the informativeness of security-related commit messages?}
\newcommand{\RqTwo}{\texorpdfstring{RQ$_2$}: Does security commit message informativeness remain consistent over temporal scope and other code-hosting platforms?}
\newcommand{\RqThree}{\texorpdfstring{RQ$_3$}: Does security commit message informativeness differ significantly by software ecosystems?}
\newcommand{\RqFour}{\texorpdfstring{RQ$_4$}: Does security commit message informativeness differ significantly among commits that comply with the Conventional Commit guidelines and others that do not?}

\newtcolorbox{summarybox}{
  colback=blue!5,   
  colframe=white,    
  boxrule=0pt,       
  arc=0pt,           
  left=0pt,
  right=0pt,
  top=3pt,
  bottom=3pt,
  before skip=3pt,
  after skip=3pt
}
\newcommand{\upar}{\,\textcolor{black!60!black}{$\mathbf{\uparrow}$}}

\newcommand{\downar}{\,\textcolor{black!80!black}{$\mathbf{\downarrow}$}}

\newcommand{\approximate}{\,\textcolor{black!80!black}{$\mathbf{\approx}$}}

\newcommand{\equal}{\,\textcolor{black!70!black}{$\mathbf{=}$}}

\begin{document}

\title[On the Informativeness of Security Commit Messages: A Large-scale Replication Study]{On the Informativeness of Security Commit Messages:\\ A Large-scale Replication Study}

\author{Syful Islam}
\email{syful.islam@telecom-paris.fr}
\orcid{0000-0002-7441-6987} 
\affiliation{\institution{LTCI, Télécom Paris, Institut Polytechnique de Paris}
  \city{Palaiseau}
  \country{France}
}

\author{Stefano Zacchiroli}
\email{stefano.zacchiroli@telecom-paris.fr}
\orcid{0000-0002-4576-136X} 
\affiliation{\institution{LTCI, Télécom Paris, Institut Polytechnique de Paris}
  \city{Palaiseau}
  \country{France}
}

\begin{abstract}
The informativeness of security-related commit messages is crucial for patch triage: when high, it enables the rapid distribution and deployment of security fixes.
Prior research (Reis et al., 2023) reported, however, that commit messages are often too uninformative to support these activities.
To assess the robustness of this negative result, we independently replicate the original study using only the information provided in the paper, without reusing any of the original artifacts (data, analysis pipeline, etc.).

Unlike the original study, we source commit data not only from GitHub, but from the entire Software Heritage archive, which includes projects hosted on many other platforms and using multiple version control systems.
We retrieve \num{50673} security-related commits and analyze their informativeness using an independent re-implementation of the techniques introduced by Reis et al.
For the same source (i.e., GitHub) and time period (from June 1999 to August 2022) as the original study, our replication confirms the original findings in a statistically significant way: security-related commit messages are, in general, not informative enough for security-focused purposes.

We then extend the original study in several ways.
Over a longer time period (from June 1999 to October 2025), we find that commit-message informativeness is worsening. 
Breaking results down by software ecosystem (Linux kernel, Ubuntu, Go, PyPI, etc.), we observe significant differences in informativeness.
Finally, we examine emerging best practices for writing commit messages, such as the Conventional Commits Specification (CCS), and again find significant differences in an unexpected direction: CCS-compliant commits are less informative than non-compliant ones.

Our findings highlight the need for cross-ecosystem analyses to understand platform- and community-specific commit-message practices, and to inform the development and adoption of universally applicable guidelines for writing informative security-related commit messages.
\end{abstract}



\maketitle
\balance
\section{Introduction} 
In today's rapidly shifting cybersecurity landscape, software applications are under constant threat of security attacks from a variety of sources and are therefore viewed as a matter of great concern~\cite{banerjee2010research}.
While many efforts have been made to prevent security attacks, they are still growing and eventually pose significant threats to IT industries worldwide.

To mitigate security attacks, developers generally propose code changes (i.e., patches) along with necessary descriptions via commit messages to the software repository, which are afterward analyzed by the maintainers.
In some cases, maintainers are hesitant to deploy updates because the proposed change is not adequately documented as security-relevant and is therefore not deemed critical~\cite{tiefenau2020security}.
Such consequences can lead to serious security attacks on target software applications.
For instance, the Equifax security breach occurred due to failure of patching a bug in Apache Struts (i.e., CVE-2017-5638) and exposed sensitive data of 143 million US consumers, yet a patch for that vulnerability had been available for months before the compromise occurred~\cite{russellequifax2017, danequifax2017}.
Hence, timely security patching is an essential safeguard and one of the most effective and widely used strategy to prevent cyberattacks on software applications~\cite{li2019keepers}.

Security patch triage management is a structured process to evaluate, prioritize, and manage security patches before they are deployed and is therefore crucial in the contemporary software development life cycle~\cite{dissanayake2022software}.
However, effective patch triage management is ultimately dependent on the description of changes via commit messages or the availability of references in public vulnerability databases~\cite{sawadogo2022sspcatcher}.
Several previous studies reported that well-structured and informative commit messages that explicitly describe security implications can facilitate automated vulnerability detection and human understanding~\cite{reis2022secom, fulton2022understanding}.
In contrast, other studies reported that commit messages often fail to provide sufficient security-related details and lack explicit references to security issues, limiting their utility for vulnerability assessment~\cite{mock2024developers, morrison2018identifying}.
Specifically, \citet{reis2023security} reported that security commit messages are not informative enough for prototyping an effective patch triage management system.

To assess the robustness of this negative result, we independently replicate the original study using only the information provided in the paper~\cite{reis2023security}, without reusing any of the original artifacts (data, analysis pipeline, etc.).
In detail, we perform a large-scale replication study on a \num{50673} security commit message dataset (from 24 June 1999 to 07 October 2025) sourced from Software Heritage to assess security commit message informativeness and whether content richness differs significantly by software ecosystem and usage of emerging guidelines for writing structured commit messages such as Conventional Commit Specification (CCS)~\cite{ccs2025}.
We used the well-known named entity recognition (NER) technique along with a pretrained spaCy model and entity dictionary, as the original work did.
Finally, we classified the information level (i.e., Excellent, Very Good, Good, Medium, Poor, and Very Poor) of security commit messages based on entity categories and predefined rules accordingly. Specifically, our work investigates the following research questions:
\begin{itemize}
    \item \textit{\RqOne}
    
    For the same source (i.e., GitHub) and time period (from June 1999 to August 2022) as the original study, our replication confirms the original findings in a statistically significant way: security-related commit messages are, in general, not informative enough for security-focused purposes.

    \item \textit{\RqTwo}

    Over a longer time period (from June 1999 to October 2025), we find that commit message informativeness is worsening.
    In addition, we find that security-related commit messages of other code-hosting platforms (e.g., GitLab, git.kernel.org, etc.) are better in term of content richness compared to GitHub.

    \item \textit{\RqThree}
    
    Security commit message informativeness differs significantly across software ecosystems, highlighting ecosystem context as an important factor in determining the documentation quality of security-related changes.

    Overall, security commit messages from operating system-related software ecosystems \syful{(e.g., Android, Linux, Ubuntu, etc.)} are generally more informative than those from other ecosystems.
 
    \item \textit{\RqFour}
    
    Perhaps surprisingly, we find that commits that adhere to emerging guidelines for writing commit messages like the Conventional Commit Specification (CCS) are \emph{less informative} than non-compliant ones.
    
    CCS guidelines alone appear insufficient to raise information levels although they intended to standardize and improve commit messages.
\end{itemize}

Overall, these findings highlight the necessity of conducting cross-ecosystem studies to better understand differences in security maintenance practices followed by stakeholders and to support the development of standard guidelines applicable to all ecosystems for writing informative security commit messages, thereby improving automated patch triage management systems. In addition, software ecosystem communities should reevaluate their existing security policies and approaches to better motivate stakeholders to implement best practices for documenting security-related changes. We believe that enriching security commit messages and maintaining this knowledge base should become a coordinated industry-wide effort, bringing benefits for IT organizations in the rapidly shifting cybersecurity landscape.

The rest of the paper is organized as follows. Section~\ref{sec:methodology} describes our replication methodology. Section~\ref{sec:results} reports our empirical study results. Section~\ref{sec:implication} discusses the implications and recommendations of our findings. Section~\ref{sec:threats} discusses the threats to validity, and Section~\ref{sec:conclusion} concludes the paper.

\textit{Data availability:} see dedicated section at the end of the paper.

\begin{figure*}
  \centering
  \includegraphics[width=0.6\linewidth]{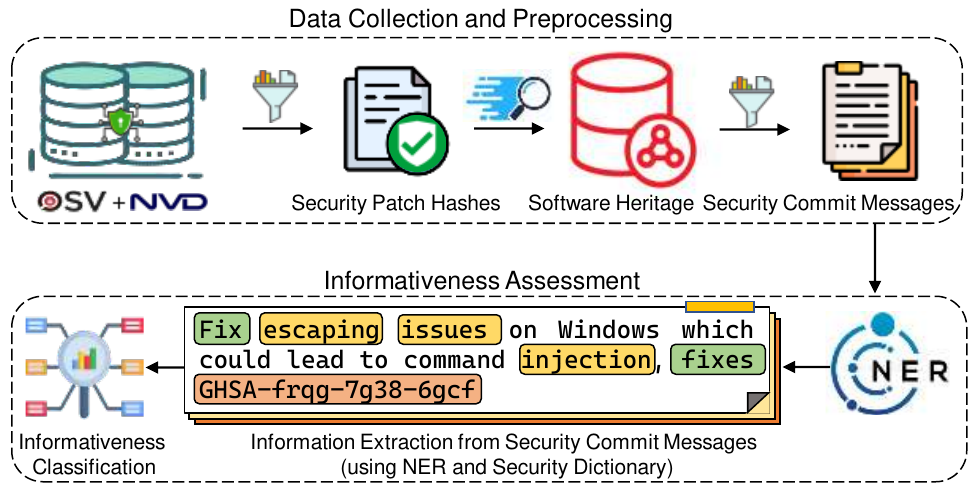}
  \caption{Overview of the methodology of our study.}
  \label{fig:method}
\end{figure*}

\section{Replication methodology}
\label{sec:methodology}
The main goal of this study is to examine the informativeness of security-related commit messages, by replicating and extending the previous study by Reis et al~\cite{reis2023security}.
We hence also follow a methodology similar to theirs, as shown in Figure~\ref{fig:method}.

\emph{Replication terminology:} In our replication of the original work we did not reuse any of the artifacts from the original paper---which does not come with a reproducibility package---data, code, etc.
What we present in the following is hence a proper \emph{replication} experiment of the original work (``Different team, different experimental setup'', according to the ACM terminology~\cite{artifactreview2025}), rather than a reproduction (``Different team, same experimental setup'').

\subsection{Data collection and preprocessing}
To build a security commit message dataset, we execute the following steps sequentially.

\subsubsection{Step 1: Vulnerability metadata collection from public vulnerability databases.}
Initially, we downloaded the latest data dump versions available online from popular open-source vulnerability publishing websites, namely OSV~\cite{osv2025} (Open-Source Vulnerability Database) and NVD~\cite{nvd2025} (National Vulnerability Database), \syful{as derived from original work}.
From OSV, we obtained a total of \num{557 427} unique vulnerability records up to 17 December 2025 for different software ecosystems such as Android, Debian, Go, Maven, and PyPI.
From NVD, we obtained a total of \num{324 423} unique CVE records up to 27 December 2025.
In total, we obtained \num{881 850} vulnerability records from both datasets.
(As vulnerability reports can be duplicated across the two data sources, we later deduplicate vulnerabilities between them, as described later in this section.)

\subsubsection{Step 2: Collection and preprocessing of references to security patches.}
Each vulnerability report from both databases (i.e., OSV and NVD) contains a ``references'' section that includes links to security patches when available.

\textit{Collection: } To collect security vulnerability records with patch references, we initially excluded both OSV and NVD records that do not contain any reference.
Afterward, we prepared a random sample from both datasets (i.e., 384 each) keeping 95\% confidence with a 5\% interval and manually analyzed the links to reveal patterns for accurate patch commit hash key (i.e., ID) extraction.
After analyzing the sample records, we gathered and merged the patterns obtained from security patch links in both datasets.
Next, we devised a regular expression based on the patterns and applied it to the links in the reference section of the vulnerability records.
\syful{Thus, we obtain a total of \num{59486} (10.67\%) unique OSV records and \num{30270} (9.33\%) unique NVD vulnerability records containing at least one security patch link.}

\textit{Deduplication and cleaning: } Since many security vulnerabilities are recorded in both NVD and OSV datasets and mentioned through the aliases field, we found duplicate patch entries between both sources.
To remove these duplicate entries and obtain unique commit hashes from vulnerability records, we executed the following operations.
First, we merged both datasets by record identifiers after necessary normalization and obtained \num{65535} unique vulnerability reports (90.77\% from OSV and 9.23\% from NVD).
When deduplicating, we kept OSV records as they provide additional useful metadata, such as ecosystem information.
Also, during deduplication, we merge missing information across the two databases; for example, if one of the two has a severity score missing from the other, we propagate it to the kept record.

Each vulnerability can be associated to multiple commits, in a one-to-many mapping, e.g., one report can contain multiple patch hashes such as GHSA-f5pm-c4cw-563p.
Hence, we extracted and merged all commit hashes referenced from all vulnerability records into a single set, de-duplicating by commit hash.
Thus, we ended up with a dataset of \num{84 435} unique security commit hashes, of which 94.07\% are from OSV and 5.93\% from the NVD dataset.

\subsubsection{Step 3: Collection and preprocessing of security commit messages from the Software Heritage archive.}
To collect the message descriptions of patched commit hashes referenced in security vulnerability records, we utilize data from the Software Heritage archive~\cite{cacm-2018-software-heritage}, the largest archive of source code and its development history, as captured by various code-hosting platforms such as Git.

We executed the following steps to obtain the final security commit message dataset.
(i) Before querying the Software Heritage graph dataset~\cite{msr-2020-challenge}, we performed exploratory analysis on the security patch commit hash list to identify anomalies and noticed that some records included short versions of hashes, whereas the database natively stores each revision under the full hash (i.e., SHA-1 full-length key) due to its global uniqueness.
In total, 581 commit hashes are short out of \num{84435} security patch hashes.
Since shortened commit hashes are not globally unique and may become ambiguous when data spans multiple repositories, leading to misidentification, we excluded short hashes with multiple returned messages to improve data accuracy and reduce false positive discovery from the Software Heritage database.
Thus, we obtained a total of \num{80 695} security patch commit messages from \num{84 435} unique hashes, corresponding to the exclusion of 65 short hashes with multiple entries and \num{3675} hashes (both short and full) that no longer exist or were missed due to a slightly earlier available version of the Software Heritage database updated up to October 2025.

(ii) We found duplicated commit messages resulting from vulnerability records with references to the same fix applied in different branches.
For example, two security commit hashes have duplicate messages because developers applied the same security fix (i.e., CVE-2015-0253) to different Apache httpd branches, so the patch was committed twice with identical descriptions as part of Apache's backporting workflow.
In such cases, since the security commit messages are the same, we kept only one of them.
Thus, an extra \num{29 367} commits were removed from the dataset, resulting in \num{51 328} security commit messages.

(iii) We then removed bot-generated commit messages and any pull request merges from Dependabot using non-human-written patterns and bot account names listed in previous works~\cite{tian2022makes, abdellatif2022bothunter, chidambaram2023dataset, golzadeh2021ground, wang2022specialized, chidambaram2025observing}.
This resulted in \num{50 978} human-written security commit messages and the removal of 350 bot/non-human-generated security commit messages.

(iv) We noticed that some of the commit messages were not written in English.
As the subsequent step used to determine the informativeness of commit messages relies on English-specific natural language processing, non-English messages should be removed to avoid underestimation.
Therefore, we ran the popular Python \texttt{langdetect}~\cite{langdetect2026} package to infer text language and detected \num{3033} (5.95\%) security commit messages not written in English.
\syful{Considering that \texttt{langdetect} package can be inaccurate when evaluating too short or too ambiguous texts, the authors manually inspected all the automatically detected non-English messages to ensure English and valid text would not be removed.}
After manual validation, we obtained a total of \num{50 673} security commit messages from 24 June 1999 to 07 October 2025 written in English, corresponding to the removal of an extra 305 non-English messages and standalone links without any text.

\subsection{Informativeness assessment}
We now detail the methodology used to extract and classify the different levels of information mentioned in security-related commit messages.

\subsubsection{Information extraction}
To accomplish this task, we employed named entity recognition (NER), which is a form of natural language processing (NLP) used to extract and identify important information from unstructured text known as entities.
The entities can be any word or bag of words that refer to the same entity category.
For instance, different forms of games such as ``Football'', ``Cricket'', and ``Golf'' are entities that belong to the same category ``Game''.
However, NER requires the design of specific entity categories and respective entity values, which relies on the target application domain for better accuracy.

In our NER design, we utilized the pretrained \texttt{en\_core\_web\_lg} NLP spaCy model, and unlike traditional rule-based approaches, no built-in linguistic rules were applied in the entity recognition process.
Instead, we customized the selected model to incorporate a predefined security-specific dictionary collected from the original study and a follow-up work by the same team~\cite{reis2023security, reis2023secomlint}.
The dictionary contains security-specific entity names and categories, which were integrated into the model as a guide for entity identification through pattern matching.

Tables~\ref{tab:entitycategory} and~\ref{tab:entityassessment} illustrate the different entity categories, rationales, and their usage in automated patch management triaging systems (i.e., detection, assessment, and prioritization), adapted from the original work.
\syful{We reused these validated entity categories and definitions to assess information richness in security commit messages.}
In summary, our NER implementation leverages NLP spaCy model flexibility while relying on the previously verified dictionary, resulting in structured entity recognition from \num{50673} security commit messages.

\begin{table}
\centering
\small
\caption{{Different Categories of Entities~\cite{reis2023security}.}}
\label{tab:entitycategory}
\begin{tabular}{@{}p{1.0 cm} p{4.0cm} p{2.5cm}@{}}
\toprule
\textbf{Category} & \textbf{Rationale} & \textbf{Examples} \\ \midrule
VULNID & Identify vulnerabilities for different ecosystems in commit messages: CVE, GHSA, OSV, PyPI, etc. This ID can enable the detection of security commits. & GHSA-269q-hmxg-m83q, CVE-2016-2512, CVE-2015-8309,  GHSA-9x4c-63pf-525f, OSV-2016-1 \\\hline
CWEID & Identify weakness type of the vulnerabilities  and can enable assessment tasks. One common taxonomy used to classify security weaknesses is the Common Weakness Enumeration (CWE) one. & CWE-119, CWE-20,  CWE-79, CWE-189 \\\hline
SEVERITY & Identify severity of discovered vulnerability through assigned score/level and can enable prioritization during patch management processes & low, medium, high, critical \\\hline
SECWORD & Security-relevant words usually describe the  vulnerability and respective fix. These words  can enable the detection of security commits. & ldap injection, crlf injection, improper validation, command injection, cross-site scripting, sanitize, bypass \\\hline
ACTION & A commit usually implies an action, in the case of security, fixing a vulnerability (corrective maintenance). & fix, patch, change, add, remove, found, protect, update, optimize, mitigate \\\hline
FLAW & Fixing a security vulnerability usually implies fixing a flaw. & defect, weakness, flaw,  fault, bug, issue \\ \bottomrule
\end{tabular}
\end{table}

\begin{table*}
\centering
\small
\caption{{Tasks in patch triage management systems (from~\cite{reis2023security}).}}
\label{tab:entityassessment}
\begin{tabular}{@{}p{2.5cm} p{7.5cm} p{4.5cm}@{}}
\toprule
\textbf{Automated Task} & \textbf{Description} & \textbf{Entity category} \\ \midrule
DETECTION (D) & Detect security-related commits through commit message analysis. & VULNID, ACTION, FLAW, SECWORD \\\hline
ASSESSMENT (A) & Classify and cluster security-related commits per weakness. & CWEID \\\hline
PRIORITIZATION (P) & Classify and order security-related commits per severity. & SEVERITY \\ \bottomrule
\end{tabular}
\end{table*}

\subsubsection{Informativeness classification}
To classify security commit message informativeness, we consider six different levels (i.e., Excellent, Very Good, Good, Medium, Poor, and Very Poor), the same used in the original work~\cite{reis2023security}.

\begin{table*}
\centering
\small
\caption{{Information spectrum of security commit message (from~\cite{reis2023security}).}}
\label{tab:assessmentinfor}
\begin{tabular}{p{1.5 cm} p{4.5cm} p{8cm} p{0.5cm} p{0.5cm} p{0.5cm}}
\toprule
\textbf{Level} & \textbf{Rule} & \textbf{Rationale} & \textbf{D} & \textbf{A} & \textbf{P} \\ \midrule
\cellcolor[HTML]{DEEAF6}Excellent & $VULNID \land CWEID \land SEVERITY \land SECWORD \land ACTION \land FLAW$ & Include all the different entity categories in a security commit message and enable all the 3 tasks (D, A, P). & \ding{51} & \ding{51} & \ding{51} \\\hline
\cellcolor[HTML]{A8D08D}Very Good & $VULNID \land SECWORD \land ACTION \land FLAW$ & Include all the different entity categories in a security commit   message except for metadata (CWEID and SEVERITY). They can still look at the   vulnerability report manually to collect the weaknesses type and severity,   but not referencing both in the commit message disables A and P. & \ding{51} &  &  \\\hline
\cellcolor[HTML]{C5E0B3}Good & $SECWORD \land ACTION \land FLAW$ & only include a description of the vulnerability and respective fix.   D is possible, but A and P will fail & \ding{51} &  &  \\\hline
\cellcolor[HTML]{FFF2CC}Medium & $ACTION \land (FLAW \lor SECWORD)$ & Include description of a flaw (that can be a security vulnerability or not) and its respective fix. D is possible but most likely will require manual validation. & \ding{51} &  &  \\\hline
\cellcolor[HTML]{FFE599}Poor & $VULNID \lor CWEID \lor SEVERITY \lor SECWORD \lor ACTION \lor FLAW$ & Include at least one of the entity categories in the security commit message. It may enable one or more of the different tasks (D, A, P), but not all. & \ding{51} & \ding{51} & \ding{51} \\\hline
\cellcolor[HTML]{F4B083}Very Poor & No entity was found. & Not included any of the entity categories. D, A, and P tasks fail. &  &  &  \\ \bottomrule
\end{tabular}
\end{table*}

Table~\ref{tab:assessmentinfor} illustrates the details about the rules based on entity categories extracted by NER, the rationales to classify the informativeness level of security commit messages, and the automated patch management tasks that could be performed.

Information levels of security commit messages are then calculated based on the presence or absence of specific entity categories and rules---again, as it was the case in the original work.

\section{Results}
\label{sec:results}
In this section, we present the analysis of security commit message informativeness levels to answer our research questions.

\subsection{\RqOne}
\textit{Motivation: } The objective of this question is to confirm the results of the original work for the same code-hosting platform and time period as described.

\textit{Approach: } To answer RQ$_1$, we compare the information level of security commit messages between the original results and our replicated-version.
Specifically, we examine the consistency of information distribution between the reported results and our replication using security commit messages from the same source (i.e., GitHub) and time period (i.e., from 24 June 1999 to 12 August 2022), as described in the original work.
This comparison allows us to assess the validity of the original results, via independent replication (i.e., Replication 1).
Finally, we apply the Mann-Whitney U test to statistically compare the results in order to assess the robustness of the conclusion.

\begin{table*}
\centering
\small
\caption{Comparison of results about the informativeness of security-related commit messages across multiple experiments: original work~\cite{reis2023security}, our replication on the same time period and forge (Replication 1), our replication on the same data source but extended time period (Replication 2), our replication on other hosting platforms and extended time period (Replication 3).}
    \label{tab:resultreplication}
\begin{tabular}{
  p{1.2cm}
  >{\raggedleft\arraybackslash}p{3.4cm}
  >{\raggedleft\arraybackslash}p{3.4cm}
  >{\raggedleft\arraybackslash}p{3.4cm}
  >{\raggedleft\arraybackslash}p{3.4cm}
}
\toprule
\textbf{Level}
& \textbf{Original result} [Source: GitHub, and Timeline: until 12-08-2022]
& \textbf{Replication 1} [Source: GitHub, and Timeline: until 12-08-2022]
& \textbf{Replication 2} [Source: GitHub, and Timeline: until 07-10-2025]
& \textbf{Replication 3} [Source: Other hosting platforms and Timeline: until 07-10-2025] \\ \midrule
\cellcolor[HTML]{DEEAF6}Excellent & \num{0}   (0.00\%) & \num{0}   (0.00\%) & \num{0}   (0.00\%) & \num{0}   (0.00\%) \\\hline
\cellcolor[HTML]{A8D08D}Very Good & \num{264}   (2.39\%)  & \num{310}   (2.09\%) & \num{523}   (2.11\%) & \num{197} (0.76\%) \\\hline
\cellcolor[HTML]{C5E0B3}Good & \num{1262}   (11.44\%) & \num{1461}   (9.86\%) & \num{2035}   (8.22\%) & \num{9267} (35.77\%) \\\hline
\cellcolor[HTML]{FFF2CC}Medium & \num{3253}   (29.48\%) & \num{4811}   (32.45\%) & \num{7685}   (31.03\%) & \num{10804} (41.70\%) \\\hline
\cellcolor[HTML]{FFE599}Poor & \num{4602}   (41.70\%) & \num{6155}   (41.52\%) & \num{10667}   (43.07\%) & \num{5158} (19.91\%) \\\hline
\cellcolor[HTML]{F4B083}Very   Poor & \num{1655}   (15.00\%) & \num{2087}   (14.08\%) & \num{3854}   (15.56\%) & \num{483} (1.86\%) \\\hline
Total & \num{11036}   (100\%) & \num{14824}   (100\%) & \num{24764}   (100\%) & \num{25909} (100\%) \\ \bottomrule
\end{tabular}
\end{table*}

\textit{Results: }
For security-related commit messages from the same source (i.e., GitHub) and time span, our replication confirms the results of the original work across all information levels as illustrated in Table~\ref{tab:resultreplication}, with no statistically significant differences observed (p-value 0.4694 > 0.05).
This reconfirms that security-related commit messages are, in general, not informative enough for security-focused purposes such as patch triage management.

In particular, the proportions of messages classified as ``excellent'' ($\Delta$=0.00\%), ``very good'' ($\Delta$=$\pm$0.30\%), ``poor'' ($\Delta$=$\pm$0.18\%), and very poor ($\Delta$=$\pm$0.92\%)  closely matched with those reported in the original work, indicating a high degree of replicability.
However, we observed some minor deviations in the ``good'' ($\Delta$=$\pm$1.58\%), and ``medium'' ($\Delta$=$\pm$2.97\%) categories, which can be attributed to differences in dataset size, pre-processing, and NER implementation.

\begin{summarybox}
\textit{\textbf{Summary of RQ$_1$:} Security-related commit messages are, in general, not informative enough for security-focused purposes when considering the same code-hosting platform (i.e., GitHub) and time period.
We confirm, We were able to replicate the original conclusions about the informativeness of security-related GitHub commits from 24 June 1999 to 12 August 2022 with no statistically significant difference.}
\end{summarybox}

\subsection{\RqTwo}
\textit{Motivation: } The objective of this question is to confirm  whether the informativeness of security-related commit message remain consistent across extended periods of time (in particular: \emph{after} the period observed in the original study) and across different code-hosting platforms (in particular: elsewhere than GitHub).

\textit{Approach: } To answer RQ$_2$, we compare the information level of security-related commit messages between the original results and our replication \syful{for longer time horizon and other code-hosting platforms.}
First, we examine the consistency between the reported results and our replication using security commit messages from the same source (i.e., GitHub) and extended time period (i.e., from 24 June 1999 to 7 October 2025).
This comparison allows us to assess whether results remain consistent for the same source and longer time horizon \syful{(i.e., Replication 2)}.

Next, we evaluate results \syful{(i.e., Replication 3)} on the extended security commit message dataset constructed from other code-hosting platforms (such as GitLab, git.kernel.org, etc.).

Finally, we apply the Mann-Whitney U test to statistically compare these findings in order to assess the robustness for security commit messages sourced from other code-hosting platforms and temporal scope.

\textit{Results:}
For security commit messages from the same source (i.e., GitHub) and extended time span, our replication finds that the informativeness of security-related commit message is \emph{worsening} over time as illustrated in Table~\ref{tab:resultreplication}, with statistically significant differences observed (p < 0.05).
In particular, we observed a moderate increase in the proportion of ``medium'' ($\Delta$=$\pm$1.55\%) and ``poor'' ($\Delta$=$\pm$1.37\%) messages, accompanied by a considerable decrease of ``good'' ($\Delta$=$\pm$3.22\%) messages.

In addition, results derived from other code hosting platforms (e.g., GitLab, git.kernel.org, etc.) show statistically significant differences (p < 0.05) across information levels compared to the original study, while the proportion of ``excellent'' security commit messages remains the same (i.e., 0.00\%).
In particular, security commit messages from other code-hosting platforms (such as git.kernel.org) are more frequently classified as ``good'' ($\Delta$=$\pm$24.33\%) and ``medium'' ($\Delta$=$\pm$12.22\%), with few messages falling under the ``very poor'' (1.86\% only and $\Delta$=$+$13.14\%) information category compared to messages sourced from GitHub.
This finding indicates that security commit messages from other code-hosting platforms are better in terms of content richness compared to GitHub.

\begin{summarybox}
\textit{\textbf{Summary of RQ$_2$:} Over a longer time period, we find that commit-message informativeness is worsening. In addition, we find that security-related commit messages on code-hosting platforms other than GitHub are better in term of informativeness compared to those found on GitHub.}
\end{summarybox}

\subsection{\RqThree}
\textit{Motivation: } The objective of this question is to verify whether security-related commit message informativeness differs significantly across software ecosystems.

\textit{Approach: } To answer RQ$_3$, \syful{first, we map ecosystem names and security commit messages using the commit hashes and software ecosystems specified in OSV vulnerability records} (e.g., Android, Ubuntu, Crates, etc.---for a full list of OSV.dev software ecosystems, see the homepage of OSV~\cite{osv2025}).
We subsequently removed vulnerability records lacking ecosystem names as references and standardized the remaining entries by eliminating any extraneous information \syful{(for example, the ecosystem name ``Ubuntu:16.04:LTS'' referenced in OSV record UBUNTU-CVE-2024-26919 is simplified to ``Ubuntu'')} to ensure consistency.
Next, we group security commit messages by ecosystem and compute the proportion of messages at each information level for ecosystems, while excluding ecosystems with fewer than 100 records from the analysis to avoid noise due to statistical measures computed on very small sets.

Finally, we apply the Kruskal-Wallis test to evaluate whether statistically significant differences exist among filtered ecosystems in commit message informativeness.
Additionally, we compute mean scores of information levels for each software ecosystem in order to rank them according to the amount of security-related information in commit messages.

\begin{table*}
\centering
\footnotesize
\caption{{Comparison of information level assessment of security commit message based on ecosystem.}}
\label{tab:levelVsecosystem}
\begin{adjustbox}{width=\textwidth}
\begin{tabular}{lrrrrrrrrrrrr@{}}
\toprule
Ecosystem/ & \textbf{Linux} & \textbf{Android} & \textbf{Ubuntu} & \textbf{Bitnami} & \textbf{Go} & \textbf{Maven} & \textbf{NuGet} & \textbf{Packagist} & \textbf{PyPI} & \textbf{RubyGems} & \textbf{Crates.io} & \textbf{npm} \\
Level &  &  &  &  &  &  &  &  &  &  &  & \\
\midrule
\cellcolor[HTML]{DEEAF6}Excellent & \num{0} (0.00\%) & \num{0} (0.00\%) & \num{0} (0.00\%) & \num{0} (0.00\%) & \num{0} (0.00\%) & \num{0} (0.00\%) & \num{0} (0.00\%) & \num{0} (0.00\%) & \num{0} (0.00\%) & \num{0} (0.00\%) & \num{0} (0.00\%) & \num{0} (0.00\%) \\\hline
\cellcolor[HTML]{A8D08D}Very   Good & \num{27} (0.82\%) & \num{9} (0.43\%) & \num{194} (0.92\%) & \num{5} (2.19\%) & \cellcolor[HTML]{A8D08D}\num{30} (2.94\%) & \num{17} (0.6\%) & \num{6} (2.2\%) & \num{16} (1.0\%) & \cellcolor[HTML]{A8D08D}\num{56} (3.09\%) & \cellcolor[HTML]{A8D08D}\num{8} (2.33\%) & \num{1} (0.32\%) & \num{10} (1.02\%) \\\hline
\cellcolor[HTML]{C5E0B3}Good & \cellcolor[HTML]{C5E0B3}\num{1283} (38.91\%) & \cellcolor[HTML]{C5E0B3}\num{815} (39.07\%) & \cellcolor[HTML]{C5E0B3}\num{6910} (32.66\%) & \num{36} (15.79\%) & \num{61} (5.98\%) & \num{94} (3.32\%) & \num{7} (2.56\%) & \num{117} (7.33\%) & \num{276} (15.25\%) & \num{15} (4.37\%) & \num{16} (5.19\%) & \num{70} (7.13\%) \\\hline
\cellcolor[HTML]{FFF2CC}Medium & \cellcolor[HTML]{FFF2CC}\num{1362} (41.31\%) & \cellcolor[HTML]{FFF2CC}\num{991} (47.51\%) & \cellcolor[HTML]{FFF2CC}\num{8521} (40.27\%) & \num{86} (37.72\%) & \num{275} (26.96\%) & \num{550} (19.44\%) & \num{51} (18.68\%) & \num{507} (31.75\%) & \num{667} (36.85\%) & \num{116} (33.82\%) & \num{83} (26.95\%) & \num{327} (33.3\%) \\\hline
\cellcolor[HTML]{FFE599}Poor & \num{599} (18.17\%) & \num{260} (12.46\%) & \num{4786} (22.62\%) & \num{82} (35.96\%) & \cellcolor[HTML]{FFE599}\num{476} (46.67\%) & \cellcolor[HTML]{FFE599}\num{1553} (54.9\%) & \cellcolor[HTML]{FFE599}\num{190} (69.6\%) & \num{629} (39.39\%) & \num{625} (34.53\%) & \num{153} (44.61\%) & \num{138} (44.81\%) & \num{426} (43.38\%) \\\hline
\cellcolor[HTML]{F4B083}Very   Poor & \num{26} (0.79\%) & \num{11} (0.53\%) & \num{747} (3.53\%) & \num{19} (8.33\%) & \num{178} (17.45\%) & \cellcolor[HTML]{F4B083}\num{615} (21.74\%) & \num{19} (6.96\%) & \cellcolor[HTML]{F4B083}\num{328} (20.54\%) & \num{186} (10.28\%) & \num{51} (14.87\%) & \cellcolor[HTML]{F4B083}\num{70} (22.73\%) & \num{149} (15.17\%) \\\hline
Total & \num{3297} (100.0\%) & \num{2086} (100.0\%) & \num{21158} (100.0\%) & \num{228} (100.0\%) & \num{1020} (100.0\%) & \num{2829} (100.0\%) & \num{273} (100.0\%) & \num{1597} (100.0\%) & \num{1810} (100.0\%) & \num{343} (100.0\%) & \num{308} (100.0\%) & \num{982} (100.0\%) \\ \bottomrule
\end{tabular}
\end{adjustbox}
\end{table*}

\textit{Results: } 
Our investigation highlights that security commit message informativeness differs significantly (p < 0.05) \syful{across the remaining software ecosystems (Linux kernel, Ubuntu, Go, PyPI, etc.) after filtering} as summarized in Table~\ref{tab:levelVsecosystem}.

Some ecosystems produce notable portion of good and medium quality messages, while others tend to have more poor and very poor quality descriptions.
According to the information level (i.e., mean scores)  ranking, operating system-related ecosystems (such as Android, Ubuntu, Linux) demonstrate better security-related commit message informativeness than other ecosystems. In particular, operating system-related ecosystems produce a high proportion of good (32.66\% -- 39.07\%) and medium (40.27\% -- 47.51\%) quality messages, with few messages falling under the very poor (0.53\% -- 3.53\%) information level category. In contrast, other ecosystems (i.e., such as Go, Maven, PyPI) produce more poor (34.53\% -- 69.6\%) and very  poor (8.33\% -- 22.73\%) quality messages; however, a few ecosystems (including Go, PyPI, RubyGems, but not all) produce a slightly higher proportion of very good messages (1.02\% -- 3.09\%) than operating system-related ecosystems.

This finding suggests that software projects with long development histories and more formal contribution processes emphasizing clarity and traceability possibly incentivize developers to write more informative security-related commit messages.


\begin{summarybox}
\textit{\textbf{Summary of RQ$_3$:} Commit message informativeness differs significantly across software ecosystems, highlighting ecosystem context as an important factor in determining documentation quality of security-related changes. Overall, security commit messages from operating system-related software ecosystems (e.g., Android, Linux, Ubuntu, etc.) are generally more informative than those from other ecosystems.}
\end{summarybox}

\subsection{\RqFour}
\textit{Motivation: } The objective of this question is to verify whether commit message informativeness differs significantly between commits that adhere to the emerging guideline for writing commit messages Conventional Commit Specification (CCS) and commits that do not.

\textit{Approach: } To answer RQ$_4$, we started by classifying the type of security commit messages into CCS-compliant and non-CCS compliant, using the popularly known Python package conventional\_pre\_commit~\cite{precommit2026}, in its default configuration (i.e., without customizing the recognized commit types).

This resulted in the identification of a total of 1850 (3.65\%) CCS-compliant and \num{48823} (96.35\%) non-CCS-compliant messages.
Afterward, we group security commit messages by CCS and non-CCS type and summarized their proportion of classified information level categories.

Finally, we conduct the Mann-Whitney U test to determine whether statistically significant differences exist in commit message informativeness between CCS-compliant and non-CCS-compliant messages.

\begin{table}
\centering
\small
\caption{{Comparison of information level between CCS (Conventional Commit Specification) and non-CCS compliant security commit message.}}
\label{tab:resultreplicationccs}
\begin{tabular}{
  p{1.5cm}
  >{\raggedleft\arraybackslash}p{2.0cm}
  >{\raggedleft\arraybackslash}p{2.0cm}
}
\toprule
\textbf{Level} & \textbf{CCS} & \textbf{Non-CCS} \\ \midrule
\cellcolor[HTML]{DEEAF6}Excellent & \num{0} (0.00\%) \equal & \num{0} (0.00\%) \equal \\\hline
\cellcolor[HTML]{A8D08D}Very   Good & \num{11} (0.59\%) \downar  & \num{709} (1.45\%) \upar \\\hline
\cellcolor[HTML]{C5E0B3}Good & \num{146} (7.89\%) \downar & \num{11156} (22.85\%) \upar \\\hline
\cellcolor[HTML]{FFF2CC}Medium & \num{686} (37.08\%) \approximate & \num{17803} (36.46\%) \approximate \\\hline
\cellcolor[HTML]{FFE599}Poor & \num{920} (49.73\%) \upar & \num{14905} (30.53\%) \downar \\\hline
\cellcolor[HTML]{F4B083}Very   Poor & \num{87} (4.70\%) \downar & \num{4250} (8.70\%) \upar \\\hline
Total & \num{1850} (100\%) \equal & \num{48823} (100\%) \equal \\ \bottomrule
\end{tabular}
\end{table}

\textit{Results: } The results summarized in Table~\ref{tab:resultreplicationccs} indicate that CCS-compliant security commit messages are \emph{less informative} compared to non-compliant ones, with statistically significant differences observed overall (p < 0.05).
In particular, CCS-compliant messages are more frequently classified as ``medium'' ($\Delta$=$\pm$0.62\%) and ``poor'' ($\Delta$=$\pm$19.20\%) quality.
In contrast, non-CCS messages exhibit a higher proportion of ``very good'' ($\Delta$=$\pm$0.86\%) and ``good'' ($\Delta$=$\pm$14.96\%) information levels.

However, a smaller portion of CCS-compliant messages fall into the ``very poor'' (4.70\%) information category compared to non-CCS security commit messages (8.70\%), indicating the necessity of explicit rule enforcement to document any security-related changes and raising awareness about the benefits of content richness among developers.


\begin{summarybox}
\textit{\textbf{Summary of RQ$_4$:} CCS-compliant security commit messages are less informative compared to non-compliant ones, indicating that CCS guidelines alone appear insufficient to raise information levels although they are intended to standardize and improve the quality and usefulness of commit messages.}
\end{summarybox}

\section{Implications and recommendations}
\label{sec:implication}

Our analysis strengthen previous results reporting that security commit messages are not informative enough for security-focused purposes, such as prototyping an effective patch triage management system.

Further analysis demonstrates that ecosystem context and code-hosting platforms are important factors in the informativeness of security-related commit messages, with statistically-significant variations across different platforms.

Moreover, we observed that emerging guidelines for writing ``better'' commit messages like Conventional Commits are insufficient to raise information levels, at least for security-related purposes like those analyzed in this study.
These findings highlight the necessity of conducting cross-software ecosystem studies to better understand differences in existing practices and to support the development of standard guidelines applicable to all ecosystems for writing security commit messages with information richness.

In summary, it is still necessary to ensure security commit messages are detailed, well-organized, and clearly explain the context of the code changes.
Below, we present the key recommendations obtained from the analysis results, aligning with previous studies, for practitioners (i.e., developers, maintainers), researchers, and educators to improve security commit message informativeness and ultimately better serve security-relevant purposes.

\subsection{Practitioners}
We recommend that practitioners follow strict rules and lexicons when writing commit messages, ensuring semantic alignment with code changes~\cite{guo2023study, zeng2021deep}.
For instance, a univocal language can be used to identify the type of operation committed by any author, such as security fix, refactor, etc.~\cite{guo2023study, zeng2021deep}.
Such semantic alignment will help other practitioners understand and extract necessary information for efficient patch management.

To improve automated security patch identification, practitioners should also avoid explicitly hiding the nature of security vulnerabilities in commit messages~\cite{nguyen2022hermes}.

Besides, messages should clearly describe the type of security issue addressed by including references (such as CVE or bug IDs) to facilitate easier linking of fixes~\cite{zuo2023commit, wang2022vcmatch, tan2021locating}.
In addition, ~\citet{morrison2018identifying} suggested that combining standard security keywords (e.g., ``encryption,'' ``authentication'') with project-specific vocabulary further improves precision in identifying security-relevant changes.
Furthermore, keeping commit messages concise with separate commit-info, commit-subject, and commit-body can improve automated understanding and accuracy of patch identification~\cite{zhou2021spi}.
Additionally, senior authors (e.g., maintainers) of a project should be proactive in reviewing commits made by new authors carefully and ensure that code changes are accompanied by well-structured and detailed messages~\cite{suzuki2017application}.

In summary, clear understanding of these recommendations by practitioners can help them gain the necessary technical knowledge to write better security commit messages, thereby better serving security-related purposes in a rapidly shifting cybersecurity landscape.

\subsection{Researchers}
Researchers should investigate ways to standardize templates and formal guidelines for writing security commit messages by analyzing existing software ecosystem practices.
They should encourage practitioners to adopt best development practices for security by highlighting how informative commit messages with structured and consistent formatting can facilitate effective patch triage management systems~\cite{reis2023security}.

Moreover, researchers should investigate ways to reduce manual verification effort, as previous studies further suggest that even when automated classifiers identify security commits, manual expert verification remains essential for ensuring precision~\cite{sabetta2018practical}.
For instance, our NER implementation requires building a security-specific vocabulary dictionary for better entity extraction and information level classification.
Besides, building a better entity dictionary relies on expert domain knowledge.
Therefore, researchers should investigate ways to build benchmark security dataset and  dictionary \syful{for better coverage of security name entities}.

\subsection{Educators}
Our findings can be leveraged by educators as a road map to design their full-stack software development courses for target ecosystems.
They should emphasize teaching the purpose and impact of writing high-quality commit messages in distributed version control systems like Git.

In addition, instructors can arrange periodical workshops for practical exercises required to write, review, and revise commit messages for reinforcing industry standard practices.
Thus, using peer review and continuous instructor feedback on commit messages can significantly improve students' awareness and writing practices.

Moreover, educators can introduce students to the several existing official conventions for writing commit messages and discuss best practices to be followed during the software development phase.

\section{Threats to validity}
\label{sec:threats}

\subsection{Internal validity}

Internal validity is threatened by the lack of access to the original reproducibility package and implementation details, which makes exact reproducibility of the methodology infeasible.
Therefore, we \emph{replicated} the methodology of the original study, based only on its description in the origin paper.
However, we communicated via email with the first author of the original work and sought guidance on the NER implementation and entity dictionary, which helped us reduce ambiguity, but full equivalence with the original implementation is not guaranteed.

We classified CCS and non-CCS security commit messages based on the predefined hooks available in the \texttt{conventional\_pre\_commit} package, which may cause CCS-compliant messages with custom hooks to be missed.
We do not see this as serious threat because, if anything, for security purposes we need general guidelines that allow to extract security-relevant information, independently from project-specific customs.
Our results provide a first empirical verification of the fact that we are not there yet.

In addition, to extract security commit hashes, we performed manual analysis of sample vulnerability records to formulate better regular expressions, which may also introduce selection bias and consequently influence the final results.

\subsection{Construct validity}

Construct validity is threatened by dataset reconstruction.
Since we did not find data extraction scripts, the constructs had to be redefined using newly curated data and custom extraction techniques.

In addition, to collect the security commit hashes, we considered commit links mentioned in vulnerability record to systematically be references to vulnerability-\emph{fixing} commits, whereas in some cases they may point to vulnerability-\emph{inducing} commits, for documentation purposes.
We share this limitation with the original study, which hence does not invalidate the usefulness of the replication part of this study.

Even with guidance from the original authors about the NER entity dictionary, the developed regular expressions may not fully capture the same constructs intended in the original study.

\subsection{External validity}

External validity is threatened by the observed differences in results when applying the methodology to different data sources.
While nearly identical results were obtained on the same data source, inconsistencies across different code-hosting platforms may occur due to data-source dependency.

\section{Conclusion}
\label{sec:conclusion}

In this paper, we independently replicate a prior original work (Reis et al., 2023) on security commit message informativeness using the only information provided, without reusing any of the original artifacts (data, analysis pipeline, etc.) to assess the robustness of their negative result as reported. In detail, we perform a large scale replication study on \num{50673} security commit message dataset (from 24 June 1999 to 07 October 2025) sourced from Software Heritage to assess security commit message informativeness and  whether content richness significantly differ by software ecosystem and standard CCS official guideline. 

Our replication confirms the original findings in a statistically significant way: security-related commit messages are, in general, not informative enough for security-focused purposes. We also find that commit-message informativeness is worsening over time; while considering GitHub. 
Further, breaking results down by software ecosystem (Linux kernel, Ubuntu, Go, PyPI, etc.), we observe significant differences in informativeness. In addition, CCS guidelines alone appear insufficient to raise informativeness of security commit message. Based on the findings, we provide recommendations for practitioners, researchers,  as well
as educators.

This study provides motivation to develop strategies for improving security commit messages. We believe that enriching security commit messages and maintaining this knowledge base should become a coordinated industry-wide effort, bringing benefits for IT organizations in rapidly shifting cybersecurity landscape.

\section*{Data availability}
A publicly available reproducibility package~\cite{replication2026} containing the curated vulnerability records with patch references, the associated security commit message dataset from the Software Heritage database, a summary of results, and all scripts used in this study are included in the package.

\begin{acks}
This work was supported by France Agence Nationale de la Recherche (ANR), program France 2030, reference ANR-22-PTCC-0001.
\end{acks}

\balance

\end{document}